\documentclass[12pt]{article}
\usepackage{latexsym}
\usepackage{cite}

\begin{document}

{\Large \bf Self-dual gravity is completely integrable} \\[4mm]
{\large \bf Y Nutku$^1$, M B Sheftel$^2$, J Kalayci$^1$ and D
Yaz{\i}c{\i}$^3$}
 \\[3mm] $^1$ Feza G\"{u}rsey Institute, PO Box 6, Cengelkoy, 81220
 Istanbul, Turkey \\
 $^2$ Department of Physics, Bo\u{g}azi\c{c}i University,
  34342 Bebek, Istanbul, Turkey \\
 $^3$ Department of Physics, Y{\i}ld{\i}z Technical
University, Esenler, Istanbul \\ \phantom{st}34210,Turkey
\vspace{1mm}
\\ E-mail:
 nutku@gursey.gov.tr,
 mikhail.sheftel@boun.edu.tr,\\ \phantom{E-mail:}
 jankalayci@gmail.com,
  yazici@yildiz.edu.tr
\begin{abstract}\noindent
 We discover multi-Hamiltonian structure of complex Monge-Amp\`ere
equation ($CMA$) set in a real first-order two-component form.
Therefore, by Magri's theorem this is a completely integrable
system in four real dimensions. We start with Lagrangian and
Hamiltonian densities and obtain a symplectic form and the
Hamiltonian operator that determines the Dirac bracket. We have
calculated all point symmetries of two-component $CMA$ system and
Hamiltonians of the symmetry flows. We have found two new real
recursion operators for symmetries which commute with the operator
of a symmetry condition on solutions of the $CMA$ system. These
operators form two Lax pairs for the two-component system. The
recursion operators, applied to the first Hamiltonian operator,
generate infinitely many real Hamiltonian structures. We show how
to construct an infinite hierarchy of higher commuting flows
together with the corresponding infinite chain of their
Hamiltonians.
\end{abstract}
 PACS numbers: 04.20.Jb, 02.40.Ky \\
AMS classification scheme numbers: 35Q75, 83C15

\section{Introduction}

In earlier paper \cite{nns} we presented complex multi-Hamiltonian
structure of Pleba\~nski's second heavenly equation \cite{pleb},
which by Magri's theorem \cite{magri} proves that it is a
completely integrable system in four complex dimensions. We expect
that Pleba\~nski's first heavenly equation also admits
multi-Hamiltonian structure, since these two equations, governing
Ricci-flat metrics with (anti-)self-dual Riemann curvature
$2$-form, are related by Legendre transformation of corresponding
heavenly tetrads \cite{pleb}. However, since both Pleba\~nski's
equations are complex, their solutions are potentials of the
complex metrics that satisfy Einstein equations in complex
four-dimensional spaces. In the case of complex Monge-Amp\`ere
equation ($CMA$), that governs (anti-)self-dual gravity in real
four-dimensional spaces with either Euclidean or ultra-hyperbolic
signature, we have an additional condition that symplectic,
Hamiltonian, and recursion operators all should be real.
Furthermore, the transformation between the two heavenly equations
cannot be applied to transform the second heavenly equation to
$CMA$ because the latter equation is real and therefore the
multi-Hamiltonian structure of $CMA$ cannot be obtained by
transforming the multi-Hamiltonian structure of the second
heavenly equation given in \cite{nns}. Therefore, in this paper we
analyze the complex Monge-Amp\`ere equation independently of our
previous work and obtain real recursion operators for symmetries
and real multi-Hamiltonian structures of $CMA$.

In section \ref{sec-1storder} we start with the complex
Monge-Amp\`ere equation in a two-component first-order
evolutionary form with the Lagrangian that is appropriate for
Hamiltonian formulation. In section \ref{sec-Hamilton} we present
a symplectic structure and Hamiltonian structure of this $CMA$
system. In section \ref{sec-real} we transform the Hamiltonian
density and Hamiltonian operator to real variables and introduce a
convenient notation needed later to arrive at a compact form of
recursion operators and higher Hamiltonian operators. In section
\ref{sec-symmetry} we derive a symmetry condition, that determines
symmetries of $CMA$ system, in a two-component form and real
variables, using our new notation. We have calculated all point
symmetries of the $CMA$ system and Hamiltonians of the symmetry
flows that yield conservation laws for the $CMA$ system. In
section \ref{sec-recursion} we obtain two new real recursion
operators for symmetries which commute with an operator of the
symmetry condition on solutions of the $CMA$ system. Moreover,
these two couples of operators form two Lax pairs for the
two-component system. The recursion operators, being applied to
the first Hamiltonian operator, yield further Hamiltonian
structures and bi-Hamiltonian reresentations of the $CMA$ system.
Repeating this procedure for the second Hamiltonian operator, we
could generate infinitely many Hamiltonian structures of the $CMA$
system. This multi-Hamiltonian structure of the $CMA$ system
proves its complete integrability in the sense of Magri and hence
complete integrability of the (anti-)self-dual gravity in four
real dimensions with either Euclidean or ultra-hyperbolic
signature. In section \ref{sec-high} we construct an infinite
hierarchy of higher flows and show the way of calculating a
corresponding infinite chain of higher Hamiltonians.

\section{Complex Monge-Amp\`ere equation in first-order evolutionary form and its Lagrangian}
\setcounter{equation}{0}
 \label{sec-1storder}

Four-dimensional hyper-K\"ahler metrics
\begin{equation}\label{metr}
  {\rm d} s^2 = u_{1\bar 1} {\rm d} z^1 {\rm d}\bar z^1 + u_{1\bar 2} {\rm d} z^1 {\rm d}\bar z^2 +
  u_{2\bar 1} {\rm d} z^2 {\rm d}\bar z^1 + u_{2\bar 2} {\rm d} z^2 {\rm d}\bar z^2
\end{equation}
satisfy Einstein field equations with either Euclidean or
ultra-hyperbolic signature, if the K\"ahler potential $u$
satisfies elliptic or hyperbolic complex Monge-Amp\`ere equation
\begin{equation}
u_{1\bar 1}u_{2\bar 2} - u_{1\bar 2}u_{2\bar 1} = \varepsilon
 \label{cma}
\end{equation}
with $\varepsilon = \pm 1$ respectively \cite{pleb}. Here $u$ is a
real-valued function of the two complex variables $z^1 ,z^2$ and
their conjugates $\bar z^1 ,\bar z^2$, the subscripts denoting
partial derivatives with respect to these variables. Such metrics
are Ricci-flat and have (anti-)self-dual curvature.

In order to discuss the Hamiltonian structure of $CMA$
(\ref{cma}), we shall replace the complex conjugate pair of
variables $z^1, \bar z^1$ by the real time variable $t=2\Re{z^1}$
and the real space variable $x=2\Im{z^1}$ and change the notation
for the second complex variable $z^2=w$. Then (\ref{cma}) becomes
\begin{equation}
(u_{tt}+u_{xx})u_{w\bar w} - u_{tw}u_{t\bar w} - u_{xw}u_{x\bar w}
+ i(u_{tw}u_{x\bar w} - u_{xw}u_{t\bar w}) = \varepsilon .
 \label{cmareal}
\end{equation}
Now we can express (\ref{cmareal}) as a pair of first-order
nonlinear evolution equations by introducing an auxiliary
dependent variable $v = u_t$
\begin{equation}
\left\{
\begin{array}{l}
u_t = v \\
v_t = -u_{xx} + \frac{\textstyle 1}{\textstyle u_{w\bar
w}}\Bigl(v_wv_{\bar w} + u_{xw}u_{x\bar w} + i(v_{\bar
w}u_{xw}-v_wu_{x\bar w}) + \varepsilon\Bigr),
\end{array}
\right.
 \label{uv}
\end{equation}
so that finally (\ref{cma}) is set in a two-component form. For
the sake of brevity we shall henceforth refer to (\ref{uv}) as
$CMA$ system.

The Lagrangian density for the original form (\ref{cma}) of the
complex Monge-Amp\`ere equation was suggested in \cite{yn}
\begin{equation}
L = \frac{1}{6} [u_1u_{\bar 1}u_{2\bar 2} + u_2u_{\bar 2}u_{1\bar
1}-u_1u_{\bar 2}u_{2\bar 1} - u_2u_{\bar 1}u_{1\bar 2}] +
\varepsilon u ,
 \label{cmalagrange}
\end{equation}
but this must be cast into a form suitable for passing to a
Hamiltonian. This requires that the form of a Lagrangian should be
appropriate for applying Dirac's theory of constraints
\cite{dirac}. We choose the Lagrangian density for the first-order
$CMA$ system (\ref{uv}) to be degenerate, that is, linear in the
time derivative of unknown $u_t$ and with no $v_t$:
\begin{eqnarray}
 L & = & \frac{1}{6}\{(u_x^2 - 3v^2)u_{w\bar w} + u_wu_{\bar
w}u_{xx} - u_x(u_wu_{x\bar w} + u_{\bar w}u_{xw}) \nonumber
\\ & & \mbox{} + u_t\Bigl(2i(u_wu_{x\bar w} - u_{\bar w}u_{xw}) + 6vu_{w\bar w})\Bigr)\} +
\varepsilon u
 \label{Luv}
\end{eqnarray}
which, after substituting $v=u_t$, coincides with our original
Lagrangian (\ref{cmalagrange}) up to a total divergence.

\section{Symplectic and Hamiltonian structures}
\setcounter{equation}{0}
 \label{sec-Hamilton}

Since the Lagrangian density (\ref{Luv}) is linear in $u_t$ and
has no $v_t$, the canonical momenta
\begin{eqnarray}
& & \pi_u = \frac{\partial L}{\partial u_t} = \frac{i}{3}
(u_wu_{x\bar w} - u_{\bar w}u_{xw}) + vu_{w\bar w} \nonumber
\\ & & \pi_v = \frac{\partial L}{\partial v_t} = 0
 \label{pi_v}
\end{eqnarray}
cannot be inverted for the velocities $u_t$ and $v_t$ and so the
Lagrangian is degenerate. Therefore, according to the Dirac's
theory \cite{dirac}, we impose them as constraints
\begin{eqnarray}
\phi_u & = & \pi_u + \frac{i}{3}(u_{\bar w}u_{xw} - u_wu_{x\bar
w}) - vu_{w\bar w} = 0 \nonumber
 \\ \phi_v & = & \pi_v = 0
 \label{constraints}
\end{eqnarray}
and calculate the Poisson brackets of the constraints (more
details of the procedure were given in \cite{nns})
\begin{equation}
K_{ik} = \left[ \phi_i(x,w,\bar w) , \phi_k(x',w',\bar w') \right]
\label{kik}
\end{equation}
collecting results in a $2\times 2$ matrix form, where the
subscripts run from $1$ to $2$ with $1$ and $2$ corresponding to
$u$ and $v$ respectively. This yields the symplectic operator $K$
that is the inverse of the Hamiltonian operator $J_0$:
\begin{equation}
 K =    \left(            \begin{array}{cc}
 (v_{\bar w}-iu_{x\bar w}) D_w + (v_w+iu_{xw}) D_{\bar w} + v_{w\bar w} & - u_{w\bar w} \\
  u_{w\bar w} & 0
\end{array}   \right)
\label{kmu}
\end{equation}
as an explicitly skew-symmetric local operator. A symplectic
2-form is a volume integral $\Omega = \int\limits_{V}\omega dx dw
d\bar w$ of the density
\begin{equation}
 \omega = \frac{1}{2} \, d u^i \wedge K_{ij} \, d u^j =
\frac{1}{2} (v_{\bar w}-iu_{x\bar w}) d u \wedge d u_w +
\frac{1}{2} (v_{w}+iu_{xw}) d u \wedge d u_{\bar w} + u_{w\bar w}
dv \wedge du
 \label{defomega}
\end{equation}
where $u^1 = u$ and $u^2 = v$. In $\omega$, under the sign of the
volume integral, we can neglect all the terms that are either
total derivatives or total divergencies due to suitable boundary
conditions on the boundary surface of the volume.

For the exterior differential of this 2-form we obtain
\begin{eqnarray}
& &  d \omega = -i d u_x \wedge d u_w \wedge d u_{\bar w} =
-(i/3)\Bigl( D_x(du \wedge du_w \wedge du_{\bar w})
 \label{domega}
\\ & & \mbox{} + D_w(du_x
\wedge du \wedge du_{\bar w}) + D_{\bar w}(du_x \wedge du_w \wedge
du)\Bigr) \iff 0 \nonumber
\end{eqnarray}
that is, a total divergence which is equivalent to zero, so that
the 2-form $\Omega$ is closed and hence symplectic. The
Hamiltonian operator $J_0$ is obtained by inverting $K$ in
(\ref{kmu})
\begin{eqnarray}
\hspace{-15.8pt} & & J_0 =
 \label{J0}
  \\ \hspace{-15.8pt} & & \left(
 \begin{array}{cc}
0 & \frac{\textstyle 1}{\textstyle u_{w\bar w}}
\\[2mm] -\frac{\textstyle 1}{\textstyle u_{w\bar w}} & \frac{\textstyle v_{\bar w}-iu_{x\bar w}}{\textstyle 2u_{w\bar
w}^2} D_w + D_w \frac{\textstyle v_{\bar w}-iu_{x\bar
w}}{\textstyle 2u_{w\bar w}^2} + \frac{\textstyle
v_w+iu_{xw}}{\textstyle 2u_{w\bar w}^2} D_{\bar w} + D_{\bar w}
\frac{\textstyle v_w+iu_{xw}}{\textstyle 2u_{w\bar w}^2}
\end{array}
\right) \nonumber
\end{eqnarray}
that is explicitly skew-symmetric. It satisfies the Jacobi
identity due to (\ref{domega}).

The Hamiltonian density is
\[ H_1 = \pi_u u_t + \pi_v v_t - L \]
with the result
\begin{equation}
 H_1 = \frac{1}{6}\Bigr[(3v^2-u_x^2)u_{w\bar w} - u_wu_{\bar w}u_{xx} +
u_x(u_{\bar w}u_{xw}+u_wu_{x\bar w})\Bigl] - \varepsilon u .
 \label{H1}
\end{equation}
$CMA$ system can now be written in the Hamiltonian form with the
Hamiltonian density $H_1$ defined by (\ref{H1})
\begin{equation}
\left(
\begin{array}{c}
u_t \\ v_t
\end{array}
\right) =  J_0 \left(
\begin{array}{c}
\delta_u H_1 \\ \delta_v H_1
\end{array}
\right)
 \label{Hamilton}
\end{equation}
where $\delta_u$ and $\delta_v$ are Euler-Lagrange operators
\cite{olv} with respect to $u$ and $v$ applied to the Hamiltonian
density $H_1$ (they correspond to variational derivatives of the
Hamiltonian functional $\int\limits_V H_1 dV$).

\section{Transformation to real variables}
\setcounter{equation}{0}
 \label{sec-real}

In the case of $CMA$, that governs (anti-)self-dual gravity with
either Euclidean or ultra-hyperbolic signature, we have an
additional condition that all the objects in the theory, in
particular a recursion operator, should be real. Therefore, we
transform the Hamiltonian density together with the symplectic and
Hamiltonian operators to the real variables $y = 2\Re{w}$ and $z =
2\Im{w}$. The Hamiltonian density in the real variables becomes \[
 H_1 = \frac{1}{6}\Bigl[(3v^2-u_x^2)\Delta(u) - (u_y^2 + u_z^2)u_{xx}
 + 2u_x(u_y u_{xy} + u_z u_{xz})\Bigr] - \varepsilon u .
 \]
where $\Delta(u) = u_{yy} + u_{zz}$, which simplifies after
cancelling terms that are total derivatives to
\begin{equation}\label{H1R}
  H_1 = \frac{1}{2}\,[v^2\Delta(u) - u_{xx}(u_y^2 + u_z^2)] -
  \varepsilon u.
\end{equation}
The transformation of the Hamiltonian operator $J_0$ in (\ref{J0})
yields
\begin{equation}
 J_0 =  \left(
\begin{array}{cc}
\phantom{-} 0 & \hspace*{2mm}\frac{\textstyle 1}{\textstyle a}
\\[2mm] -\frac{\textstyle 1}{\textstyle a} & \hspace*{2mm}\frac{\textstyle 1}{\textstyle a^2}
(cD_y - bD_z) + (D_y c - D_z b)\frac{\textstyle 1}{\textstyle a^2}
\end{array}
\right) \label{J0R}
\end{equation}
where we introduce the notation
\begin{equation}\label{not}
  a = \Delta(u),\quad b = u_{xy} - v_z,\quad c = v_y +
  u_{xz},\quad Q = \frac{b^2 + c^2 + \varepsilon}{a}
\end{equation}
that we will use from now on throughout the paper, with $D_y, D_z$
designating operators of total derivatives with respect to $y, z$
respectively and $\Delta = D_y^2 + D_z^2$ is the two-dimensional
Laplace operator.

The symplectic operator (\ref{kmu}) in the real variables becomes
\begin{equation}\label{K}
  K = J_0^{-1} = \left(
  \begin{array}{cr}
  cD_y - bD_z + D_yc - D_zb & -a
  \\ a & 0
  \end{array}\right)
\end{equation}
in an explicitly skew-symmetric form.

$CMA$ system (\ref{uv}) in the real variables becomes
\begin{equation}\label{syst}
\left(
\begin{array}{c}
u_t \\ v_t
\end{array}
\right) =  J_0 \left(
\begin{array}{c}
\delta_u H_1 \\ \delta_v H_1
\end{array}
\right) = \left(
\begin{array}{c} v \\ Q - u_{xx}
\end{array}
\right)
\end{equation}
or $u_t = v, \; v_t = Q - u_{xx}$.

The four-dimensional hyper-K\"ahler metrics (\ref{metr}) in the
real variables in the notation (\ref{not}) become
\begin{equation}\label{realmetr}
 {\rm d} s^2 = \frac{1}{4}\Bigl[Q({\rm d} t^2 + {\rm d} x^2) + a({\rm
 d} y^2 + {\rm d} z^2)\Bigr] + \frac{1}{2}\Bigl[c({\rm d} t{\rm d} y + {\rm
 d} x{\rm d} z) - b({\rm d} t{\rm d} z - {\rm d} x{\rm d} y)\Bigr].
\end{equation}
The metrics (\ref{realmetr}) satisfy Einstein field equations with
either Euclidean or ultra-hyperbolic signature, if the
two-component potential $(u,v)$ in the definitions (\ref{not}) of
$a, b ,c$, and $Q$ satisfies the Hamiltonian $CMA$ system
(\ref{syst}) with $\varepsilon = +1$ or $\varepsilon = -1$
respectively . These metrics are again Ricci-flat and have
(anti-)self-dual curvature.

\section{Symmetries and integrals of motion}
\setcounter{equation}{0}
 \label{sec-symmetry}

Now, consider Lie group of transformations of the system
(\ref{syst}) in the evolutionary form, when only dependent
variables are transformed, and let $\tau$ be the group parameter.
Then Lie equations read
\begin{equation}\label{Lie}
  u_\tau = \varphi,\qquad v_\tau = \psi
\end{equation}
where $\Phi = \left(
 \begin{array}{c}
    \varphi
    \\ \psi
 \end{array}
\right)$ is a two-component symmetry characteristic of the system
(\ref{syst}). The differential compatibility conditions of
equations (\ref{syst}) and (\ref{Lie}) in the form $u_{t\tau} -
u_{\tau t} = 0$ and $v_{t\tau} - v_{\tau t} = 0$ result in the
linear matrix equation
\begin{equation}\label{symeq}
  {\cal A}(\Phi) = 0
\end{equation}
where ${\cal A}$ is the Frech\'et derivative of the flow
(\ref{syst})
\begin{equation}
 {\cal A} =  \left( \begin{array}{cc} D_t & - 1 \\ D_x^2
 - \frac{\textstyle 2}{\textstyle a}(cD_z + bD_y)D_x + \frac{\textstyle
 Q}{\textstyle a} \Delta ,
 & D_t - \frac{\textstyle 2}{\textstyle a} (cD_y - bD_z)
 \end{array}
\right)
 \label{A}
\end{equation}
where the first row of (\ref{symeq}) yields $\varphi_t = \psi$.

Using the software packages LIEPDE and CRACK by T. Wolf
\cite{wolf}, run under REDUCE 3.8, we have calculated all point
symmetries of $CMA$ system (\ref{syst}), a class of solutions of
the matrix equation (\ref{symeq}). We list their generators and
two-component symmetry characteristics \cite{olv}, the latter
denoted by $\varphi^u, \varphi^v$
\begin{eqnarray}
& & X_1 = t\partial_t + x\partial_x + u\partial_u, \quad
\varphi_1^u = u -tv - xu_x,\quad \varphi_1^v = t(u_{xx} - Q) -
xv_x
 \nonumber
\\ & & X_2 = z\partial_y - y\partial_z, \quad \varphi_2^u = yu_z -
zu_y,\quad \varphi_2^v = yv_z - zv_y
 \nonumber
  \\ & & X_3 = \partial_z, \quad \varphi_3^u = u_z,\quad
  \varphi_3^v = v_z \nonumber
    \\ & & X_4 = \partial_y, \quad \varphi_4^u = u_y,\quad
  \varphi_4^v = v_y
  \label{point}
  \\ & & X_5 = y\partial_y + z\partial_z + u\partial_u +
  v\partial_v,\quad \varphi_5^u = u - yu_y - zu_z,\; \varphi_5^v = v - yv_y - zv_z
  \nonumber
  \\ & & X_\alpha = \alpha(t,x,y,z)\partial_u +
  \alpha_t(t,x,y,z)\partial_v,\quad \varphi_\alpha^u =
  \alpha,\quad \varphi_\alpha^v = \alpha_t
  \nonumber
  \\ & & X_\beta = \beta_z(y,z)\partial_x -
  \beta_y(y,z)\partial_t,\; \varphi_\beta^u = \beta_y v -
  \beta_z u_x, \varphi_\beta^v = \beta_y (Q - u_{xx}) -
  \beta_z v_x
  \nonumber
\end{eqnarray}
where $\alpha(t,x,y,z)$ is an arbitrary smooth solution of the
equations
\begin{equation}\label{alfaeq}
  \Delta(\alpha) = 0,\quad \alpha_{tt} + \alpha_{xx} = 0,\quad
  \alpha_{tz} - \alpha_{xy} = 0,\quad \alpha_{ty} + \alpha_{xz} =
  0
\end{equation}
whereas $\beta(y,z)$ satisfies the two-dimensional Laplace
equation $\Delta(\beta) = 0$.

We shall find the integrals of motion generating the point
symmetries that serve as Hamiltonians of the symmetry flows
\begin{equation}
 \left(
 \begin{array}{c}
  u_\tau \\ v_\tau
 \end{array}
 \right) =
 \left(
 \begin{array}{c}
 \varphi_u \\ \varphi_v
 \end{array}
 \right) =  J_0 \left(
 \begin{array}{c}
 \delta_u H \\ \delta_v H
 \end{array}
 \right)
 \label{flow}
 \end{equation}
where the symmetry group parameter $\tau$ plays the role of time
for the symmetry flow (\ref{flow}) and ${\cal
H}=\int_{-\infty}^{+\infty} H dx dy dz$ is an integral of the
motion along the flow (\ref{syst}), with the conserved density
$H$, that generates the symmetry with the two-component
characteristic $\varphi_u, \varphi_v$. The second equality in
(\ref{flow}) is the Hamiltonian form of Noether's theorem that
gives a relation between symmetries and integrals.

We choose here the Poisson structure determined by our first
Hamiltonian operator $J_0$ since we know its inverse $K$ given by
(\ref{K}) which is used in the inverse Noether theorem
\begin{equation}
\left(
\begin{array}{c}
\delta_u H \\ \delta_v H
\end{array}
\right) = K \left(
\begin{array}{c}
\varphi_u \\ \varphi_v
\end{array}
\right)
 \label{noether}
\end{equation}
determining conserved densities $H$ corresponding to known
symmetry characteristics $\varphi_u, \varphi_v$.

Using (\ref{noether}), we reconstruct the Hamiltonians of the
flows (\ref{flow}) for all variational point symmetries in
(\ref{point}). For the scaling symmetries generated by $X_1$ and
$X_5$, Hamiltonians do not exist and so they are not variational
symmetries. For the rotational symmetry generated by $X_2$, the
Hamiltonian is
\begin{equation}\label{H2}
  H_2 = v(yu_z - zu_y)\Delta(u) - u_x [2(zu_y + yu_z)u_{yz} + u_y^2 +
  u_z^2].
\end{equation}
For the translational symmetries generated by $X_3$ and $X_4$, the
corresponding Hamiltonians $H_3$ and $H_4$ are
\begin{eqnarray}
 & & H_3 = vu_z\Delta(u) + \frac{2}{3}\,u_x(u_yu_{zz} - u_zu_{yz})
 \nonumber
\\ & & H_4 = vu_y\Delta(u) + \frac{2}{3}\,u_x(u_yu_{yz} -
u_zu_{yy}).
 \label{H34}
\end{eqnarray}
For the infinite Lie pseudogroups generated by $X_\alpha$ and
$X_\beta$, the Hamiltonians are
\begin{equation}\label{H_alfa}
  H_\alpha = \alpha v\Delta(u) + \frac{1}{2}\,\alpha_t(u_y^2 +
  u_z^2) + \alpha(u_zu_{xy} - u_yu_{xz})
\end{equation}
and
\begin{equation}\label{H_beta}
  H_\beta = \left(\frac{\beta_y}{2}\,v^2 -
  \beta_zu_xv\right)\Delta(u) - \frac{\beta_y}{2}\,u_{xx}(u_y^2 +
  u_z^2) + \frac{1}{2}\,u_x^2(\beta_{yy}u_y + \beta_{yz}u_z) -
  \varepsilon \beta_y u.
\end{equation}
In particular, the Hamiltonian of time translations
$X_{\beta=y}=-\partial_t$, that is $H_{\beta=y} = H_1$, coincides
with the Hamiltonian (\ref{H1R}) of $CMA$ flow. For translations
in $x$, $X_{\beta=z}=\partial_x$, the Hamiltonian is $H_{\beta=z}
= -u_xv\Delta(u)$. For a simple example of the symmetry
$X_\alpha$, $X_{\alpha=z} = z\partial_u$, the Hamiltonian is
$H_{\alpha=z} = zv\Delta(u) + u_xu_y$, which coincides with the
Hamiltonian $H_0$ (\ref{H0}) from an infinite chain of
Hamiltonians for a hierarchy of higher commuting flows in section
\ref{sec-recursion}.

All of these Hamiltonians of the symmetry flows are conserved
densities of the $CMA$ flow (\ref{syst}).

\section{Recursion operators and bi-Hamiltonian\\ representations of CMA system}
\setcounter{equation}{0}
 \label{sec-recursion}

Complex recursion operators for symmetries of the heavenly
equations of Pleba\~nski were introduced in the papers of Dunajski
and Mason \cite{dm2,dm}. We have used them in our method of
partner symmetries for obtaining non-invariant solutions of
complex Monge-Amp\`ere equation \cite{mnsbig,mns} and second
heavenly equation of Pleba\~nski \cite{mns} and, in a
two-component form, for generating multi-Hamiltonian structure of
Pleba\~nski's second heavenly equation in \cite{nns}. However, for
$CMA$ we have an additional condition that the equation and its
symmetries are real and hence recursion operators should also be
real. This condition leads to a couple of real recursion operators
in a $2\times 2$ matrix form. The first one is
\begin{eqnarray}\label{R1}
 & & R_1 = \left(
  \begin{array}{cr}
  0 & 0
  \\ QD_z - cD_x & b
  \end{array}
  \right)
  \\ & & \mbox{} +
 \Delta^{-1}\left(
  \begin{array}{cc}
 D_y \Bigl(- a D_x + bD_y + cD_z \Bigr) + D_z\Bigl(cD_y -
 bD_z\Bigr)
 & - D_z a
 \\[2mm] \begin{array}{l}
  D_x \Bigl[D_y\Bigl(cD_y - bD_z\Bigr) + D_z\Bigl(aD_x - bD_y -
 cD_z\Bigr)\Bigr]
 \end{array}
 & - D_xD_y a
  \end{array}
  \right) \nonumber
\end{eqnarray}
where $\Delta^{-1}$ means operator multiplication, and the second
recursion operator reads
\begin{eqnarray}\label{R2}
 & & R_2 = \left(
  \begin{array}{cr}
   0 & 0 \\
   bD_x - QD_y & c
  \end{array}
  \right)
  \\ & & \mbox{} + \Delta^{-1}
  \left(
  \begin{array}{lr}
 D_y(bD_z - cD_y) + D_z(-aD_x + bD_y + cD_z)
 & D_y a
  \\ D_x\Bigl[D_y(- aD_x + bD_y + cD_z) + D_z(cD_y - bD_z)\Bigr]
 & - D_xD_z a
  \end{array}
  \right) \nonumber
\end{eqnarray}

 Straightforward, though cumbersome, calculations show that the
operators $R_1$ and $R_2$ commute with the operator ${\cal A}$
(\ref{A}) of the symmetry condition (\ref{symeq}) on solutions of
equations (\ref{syst}) and therefore they are indeed recursion
operators for symmetries of the $CMA$ system. This means that if
$(\tilde\varphi, \tilde\psi)$ is obtained by transforming a
two-component symmetry characteristic $(\varphi, \psi)$ of the
system (\ref{syst}) by the operator $R_1$ or $R_2$
\begin{equation}\label{recursion}
  \left(
  \begin{array}{c}
  \tilde\varphi
  \\ \tilde\psi
  \end{array}
  \right)
  =
  R_i \left(
  \begin{array}{c}
  \varphi
  \\ \psi
  \end{array}
  \right)
\end{equation}
where $i = 1,2$, then $(\tilde\varphi, \tilde\psi)$ is also a
symmetry characteristic of (\ref{syst}).

Moreover, vanishing of the commutators $[R_i,{\cal A}]$, computed
without using the equations of motion (\ref{syst}), reproduces the
$CMA$ system (\ref{syst}) and hence the operators $R_i$ and ${\cal
A}$ form two real Lax pairs for the two-component system. Indeed,
introducing a short-hand notation $F = u_t - v$, $G = v_t + u_{xx}
- Q$, $\Phi = F_{xy} - G_z$ and $\chi = F_{xz} + G_y$, we rewrite
the $CMA$ system in the form $F = 0,\; G = 0$, so that $\Phi = 0$
and $\chi = 0$ on its solutions, and the first commutator reads
\begin{eqnarray}\label{commut}
 & & [R_1 , {\cal A}] =
 \left(
 \begin{array}{cc}
  0 & 0 \\
  \frac{\textstyle 1}{\textstyle a}\left[Q\Delta(F) - 2(b\Phi + c\chi)\right]D_z & -\Phi
 \end{array}
 \right) \mbox{} + \Delta^{-1}\times
 \\ & &  \left(
 \begin{array}{lc}
  \begin{array}{l}
 \Phi(D_z^2 - D_y^2) - 2\chi D_yD_z
 \\ \mbox{} + D_y\Delta(F)D_x - \Delta(F_x)D_y -
 \Delta(G)D_z ,
   \end{array}
 &\hspace*{5mm} D_z\Delta(F)
 \\  \begin{array}{c}
     2(D_z\chi D_x + D_x\Phi D_y)D_z - D_z\Delta(F)D_x^2
     \\ \mbox{} + \chi_x(D_z^2 - D_y^2) + [2\chi_z - \Delta(G)]D_xD_y
      \\ \mbox{} + \Delta(G_y)D_x - \Delta(G_x)D_y
     \end{array}
  \hspace*{5mm} & D_xD_y\Delta(F)
 \end{array}
  \right)
  \nonumber
\end{eqnarray}
and we have a similar expression for $[R_2, {\cal A}]$. Thus,
$[R_i, {\cal A}] = 0$ implies $F = 0,\; G = 0$, that is, the $CMA$
system (\ref{syst}).

These real Lax pairs are formed by the recursion operators for
symmetries and operator ${\cal A}$ of the symmetry condition and
so they are Lax pairs of the Olver-Ibragimov-Shabat type
\cite{olver,ibr}, which is different from the complex Lax pairs
suggested by Mason and Newman \cite{masnew,mw} and Dunajski and
Mason \cite{dm2,dm} and those that we used in \cite{mnsbig,mns} in
relation to partner symmetries, even if we set our new Lax pairs
in one-component forms. Furthermore, the commutator of the complex
recursion operator of Mason-Dunajski with the operator of the
symmetry condition, in a one-component form, reproduces the
symmetry condition and not the original equation $CMA$
\cite{mnsbig}.

By the theorem of Magri, given a Hamiltonian operator $J$ and a
recursion operator $R$, $RJ$ is also a Hamiltonian operator
\cite{magri}. Thereby, acting by the recursion operator $R_1$ on
the first Hamiltonian  operator $J_0$ (\ref{J0}), we obtain the
second Hamiltonian operator
 \begin{eqnarray}\label{J1}
& & J_1 = R_1J_0 = \Delta^{-1}\left(
\begin{array}{cc}
 D_z & -D_xD_y
\\ D_xD_y & D_x^2D_z
\end{array}
 \right) +
 \\[2mm] & & \left(\begin{array}{cc}
 \phantom{-} 0 & \frac{\textstyle b}{\textstyle a}
 \\[2mm] -\frac{\textstyle b}{\textstyle a} &
 \frac{\textstyle c}{\textstyle a^2}\Bigl(bD_y - aD_x\Bigr)
 + \Bigl(D_y b - D_x a\Bigr)\frac{\textstyle c}{\textstyle a^2}
 + \frac{\textstyle Q_-}{\textstyle 2a}D_z
 + D_z\frac{\textstyle Q_-}{\textstyle 2a}
\end{array}\right) \nonumber
 \end{eqnarray}
that is explicitly skew-symmetric. Here $Q_- = (c^2 - b^2 +
\varepsilon)/a$. The proof of the Jacobi identity for $J_1$ is
lengthy but can be somewhat facilitated by using Olver's criterion
in terms of functional multi-vectors \cite{olv}.

Similarly, acting by the recursion operator $R_2$ on the
Hamiltonian  operator $J_0$, we obtain a Hamiltonian operator that
is another companion for $J_0$:
 \begin{eqnarray}\label{J^1}
& & J^1 = R_2J_0 = \Delta^{-1}\left(
\begin{array}{cc}
 D_y & D_xD_z
\\ - D_xD_z & D_x^2D_y
\end{array}
 \right) +
 \\[2mm] & & \left(\begin{array}{cc}
 \phantom{-} 0 & - \frac{\textstyle c}{\textstyle a}
 \\[2mm] \frac{\textstyle c}{\textstyle a} &
 \frac{\textstyle b}{\textstyle a^2}\Bigl(cD_z - aD_x\Bigr)
 + \Bigl(D_z c - D_x a\Bigr)\frac{\textstyle b}{\textstyle a^2}
 + \frac{\textstyle Q^-}{\textstyle 2a}D_y
 + D_y\frac{\textstyle Q^-}{\textstyle 2a}
\end{array}\right) \nonumber
 \end{eqnarray}
that is also explicitly skew-symmetric and $Q^- = (b^2 - c^2 +
\varepsilon)/a$. The Jacobi identity for $J^1$ was also proved by
using Olver's criterion in terms of functional multi-vectors
\cite{olv}.

The flow (\ref{syst}) can be generated by the Hamiltonian operator
$J_1$ from the Hamiltonian density
\begin{equation}\label{H0}
  H_0 = zv\Delta(u) + u_xu_y
\end{equation}
so that CMA in the two-component form (\ref{syst}) admits two
Hamiltonian representations
\begin{equation}\label{biHam}
\left(
\begin{array}{c}
u_t \\ v_t
\end{array}
\right) =  J_0 \left(
\begin{array}{c}
\delta_u H_1 \\ \delta_v H_1
\end{array}
\right) =  J_1 \left(
\begin{array}{c}
\delta_u H_0 \\ \delta_v H_0
\end{array}
\right)
\end{equation}
and thus this is a {\it bi-Hamiltonian system}.

The same flow (\ref{syst}) can also be generated by the
Hamiltonian operator $J^1$ from the Hamiltonian density
\begin{equation}\label{H^0}
  H^0 = yv\Delta(u) - u_xu_z
\end{equation}
which yields another bi-Hamiltonian representation of the $CMA$
system (\ref{syst})
\begin{equation}\label{biHam2}
\left(
\begin{array}{c}
u_t \\ v_t
\end{array}
\right) =  J_0 \left(
\begin{array}{c}
\delta_u H_1 \\ \delta_v H_1
\end{array}
\right) =  J^1 \left(
\begin{array}{c}
\delta_u H^0 \\ \delta_v H^0
\end{array}
\right).
\end{equation}

Repeating this procedure $n$ times, we obtain a {\it
multi-Hamiltonian representation} of the $CMA$ system with the
Hamiltonian operators $J_n = R_1^n J_0$, $J^n = R_2^n J_0$,
$J_m^{n-m} = R_1^m R_2^{n-m}J_0$ ($m = 1, 2, \ldots n-1$) and
corresponding Hamiltonian densities. This procedure will be
considered in more detail in the next section for the operator
$R_1$. Multi-Hamiltonian structure of the $CMA$ system proves its
complete integrability in the sense of Magri and hence the
complete integrability of the (anti-)self-dual gravity in four
real dimensions with either Euclidean or ultra-hyperbolic
signature.

A totally different recursion operator for the (anti-)self-dual
gravity in complex Einstein spaces was obtained much earlier by
Strachan \cite{strach} by using a Legendre transformed version of
the first heavenly equation, that was derived by Grant
\cite{grant}. This recursion operator can be factorized which
suggests a bi-Hamiltonian structure of the resulting evolutionary
equation, though that was not completely proved. However, the
evolutionary equation and the related Hamiltonian structures are
expressed in complex variables, with the complex "time" t in
particular, and with a complex unknown. Therefore, the
corresponding metric will not correspond to anti-self-dual gravity
in real Einstein spaces with the Euclidean signature $(++++)$.
Furthermore, the Poisson bracket contains the unusual operator
$\partial_t^{-1}$ that could be avoided in a two-component
formulation.

\section{Infinite hierarchy of higher flows}
\setcounter{equation}{0}
 \label{sec-high}

The operators $J_0$ and $J_1$ are compatible Hamiltonian
operators, i.e. they form a Poisson pencil. This means that every
linear combination $C_0J_0 + C_1J_1$ with constant coefficients
$C_0$ and $C_1$ satisfies the Jacobi identity. This can be more
easily verified by using the Olver's criterion in terms of
functional multi-vectors though the calculation is still very
lengthy. We know from the work of Fuchssteiner and Fokas \cite{ff}
(see also the survey \cite{sheftel} and references therein) that
if a recursion operator has a factorized form, as in our case $R_1
= J_1J_0^{-1}\equiv J_1K$, and the factors $J_0$ and $J_1$ are
compatible Hamiltonian operators, then $R_1$ is hereditary
(Nijenhuis) recursion operator, i.e. it generates an Abelian
symmetry algebra out of commuting symmetry generators. Moreover,
Hermitian conjugate hereditary recursion operator $R_1^\dagger =
J_0^{-1}J_1 = KJ_1$, acting on the vector of variational
derivatives of an integral of the flow, yields a vector of
variational derivatives of some other integral of this flow. Then
(\ref{biHam}) implies that $R_1^\dagger$ generates the Hamiltonian
density $H_1$ from $H_0$:
\begin{equation}\label{R+}
  R_1^\dagger \left(
  \begin{array}{c}
  \delta_uH_0 \\ \delta_vH_0
  \end{array}
  \right) = J_0^{-1}J_1
  \left(
  \begin{array}{c}
  \delta_uH_0 \\ \delta_vH_0
  \end{array}
  \right) = \left(
  \begin{array}{c}
  \delta_uH_1 \\ \delta_vH_1
  \end{array}
  \right)
\end{equation}
where $R_1^\dagger$ is defined by
\begin{eqnarray}\label{R^+}
 & & R_1^\dagger = \left(
  \begin{array}{cc}
  0 & D_x c - D_z Q
  \\ 0  & b
  \end{array}
  \right)
  \\ & & \mbox{} +
 \left(
  \begin{array}{cc}
  \begin{array}{c}
 \Bigl(- D_x a + D_y b + D_z c\Bigr)D_y
 \\ + \Bigl(D_y c - D_z b\Bigr)D_z
 \end{array}
 & \begin{array}{c}
 \Bigl[\Bigl(D_z b - D_y c\Bigr)D_y +
 \\  \Bigl(- D_x a + D_y b +
 D_z c\Bigr)D_z\Bigr]D_x
 \end{array}
 \\[6mm] aD_z & a D_xD_y
  \end{array}
  \right)\Delta^{-1}. \nonumber
\end{eqnarray}

The first higher flow of the hierarchy is generated by $J_1$
acting on the vector of variational derivatives of $H_1$
\begin{equation}\label{flow1}
 \left(
\begin{array}{c}
u_{t_1} \\ v_{t_1}
\end{array}
\right) =  J_1 \left(
\begin{array}{c}
\delta_u H_1 \\ \delta_v H_1
\end{array}
\right)
\end{equation}
where $t_1$ is the time variable of the higher flow. This flow is
nonlocal and the right-hand side of (\ref{flow1}) is too lengthy
to be presented here explicitly.

Now we could generate the next Hamiltonian $H_2$ of the hierarchy
of commuting flows by applying $R_1^\dagger$ to the vector of
variational derivatives of $H_1$:
\begin{equation}\label{H_2}
  R_1^\dagger \left(
  \begin{array}{c}
  \delta_uH_1 \\ \delta_vH_1
  \end{array}
  \right) = KJ_1
  \left(
  \begin{array}{c}
  \delta_uH_1 \\ \delta_vH_1
  \end{array}
  \right) = \left(
  \begin{array}{c}
  \delta_uH_2 \\ \delta_vH_2
  \end{array}
  \right).
\end{equation}
Therefore, the second higher flow in the hierarchy has a
bi-Hamiltonian representation
\begin{equation}\label{flow2}
 \left(
\begin{array}{c}
u_{t_2} \\ v_{t_2}
\end{array}
\right) =  J_1 \left(
\begin{array}{c}
\delta_u H_2 \\ \delta_v H_2
\end{array}
\right) = J_1
 R_1^\dagger \left(
  \begin{array}{c}
  \delta_uH_1 \\ \delta_vH_1
  \end{array}
  \right) = J_2
 \left(
 \begin{array}{c}
 \delta_u H_1 \\ \delta_v H_1
 \end{array}
 \right)
\end{equation}
where the third Hamiltonian operator $J_2$ is generated by acting
with $R_1$ on $J_1$: $J_1R_1^\dagger = J_1KJ_1 = R_1J_1 = J_2$.
Acting by $J_2$ on the variational derivatives of $H_0$, we obtain
the relations
\begin{eqnarray}
  & & J_2 \left(
  \begin{array}{c}
  \delta_uH_0 \\ \delta_vH_0
  \end{array}
  \right) = J_1 R_1^\dagger \left(
  \begin{array}{c}
  \delta_uH_0 \\ \delta_vH_0
  \end{array}
  \right) = J_1 \left(
  \begin{array}{c}
  \delta_uH_1 \\ \delta_vH_1
  \end{array}
  \right)
  \nonumber
  \\ & & = J_0 R_1^\dagger \left(
  \begin{array}{c}
  \delta_uH_1 \\ \delta_vH_1
  \end{array}
  \right) = J_0 \left(
  \begin{array}{c}
  \delta_uH_2 \\ \delta_vH_2
  \end{array}
  \right)
  \label{ham3}
\end{eqnarray}
where we have used that $J_1 = J_0(KJ_1) = J_0R_1^\dagger$. From
(\ref{ham3}) we obtain three-Hamiltonian representation of the
first higher flow
\begin{equation}\label{3_ham}
  \left(
  \begin{array}{c}
  u_{t_1} \\ v_{t_1}
  \end{array}
  \right) = J_1 \left(
  \begin{array}{c}
  \delta_uH_1 \\ \delta_vH_1
  \end{array}
  \right) = J_2 \left(
  \begin{array}{c}
  \delta_uH_0 \\ \delta_vH_0
  \end{array}
  \right) = J_0 \left(
  \begin{array}{c}
  \delta_uH_2 \\ \delta_vH_2
  \end{array}
  \right).
\end{equation}

We could also construct the Hamiltonian $H_{-1}$ such that $H_0$
is generated from $H_{-1}$ by $R_1^\dagger$
\begin{equation}\label{H_{-1}}
 R_1^\dagger \left(
  \begin{array}{c}
  \delta_uH_{-1} \\ \delta_vH_{-1}
  \end{array}
  \right) = J_0^{-1}J_1
  \left(
  \begin{array}{c}
  \delta_uH_{-1} \\ \delta_vH_{-1}
  \end{array}
  \right)
  =  \left(
 \begin{array}{c}
 \delta_u H_0 \\ \delta_v H_0
 \end{array}
 \right)
\end{equation}
that implies a bi-Hamiltonian representation for the zeroth flow
\begin{equation}\label{flow0}
  \left(
  \begin{array}{c}
  u_{t_0} \\ v_{t_0}
  \end{array}
  \right) = J_0 \left(
 \begin{array}{c}
 \delta_u H_0 \\ \delta_v H_0
 \end{array}
 \right) = J_1 \left(
  \begin{array}{c}
  \delta_uH_{-1} \\ \delta_vH_{-1}
  \end{array}
  \right).
\end{equation}
Further we obtain
\begin{equation}
 J_2 \left(
  \begin{array}{c}
  \delta_uH_{-1} \\ \delta_vH_{-1}
  \end{array}
  \right) = J_1 R_1^\dagger \left(
  \begin{array}{c}
  \delta_uH_{-1} \\ \delta_vH_{-1}
  \end{array}
  \right) = J_1 \left(
  \begin{array}{c}
  \delta_uH_0 \\ \delta_vH_0
  \end{array}
  \right)
  \label{J012}
\end{equation}
and the bi-Hamiltonian representation (\ref{biHam}) of the
original two-component $CMA$ flow becomes a three-Hamiltonian
representation of this flow
\begin{equation}\label{3Ham}
\left(
\begin{array}{c}
u_t \\ v_t
\end{array}
\right) =  J_0 \left(
\begin{array}{c}
\delta_u H_1 \\ \delta_v H_1
\end{array}
\right) =  J_1 \left(
\begin{array}{c}
\delta_u H_0 \\ \delta_v H_0
\end{array}
\right) = J_2 \left(
  \begin{array}{c}
  \delta_uH_{-1} \\ \delta_vH_{-1}
  \end{array}
  \right).
\end{equation}

We could still continue by applying $R_1^\dagger$ to the vector of
variational derivatives of $H_2$ to generate the next Hamiltonian
$H_3$
\begin{equation}\label{H_3}
 R_1^\dagger \left(
  \begin{array}{c}
  \delta_uH_2 \\ \delta_vH_2
  \end{array}
  \right) = KJ_1 \left(
  \begin{array}{c}
  \delta_uH_2 \\ \delta_vH_2
  \end{array}
  \right) = \left(
  \begin{array}{c}
  \delta_uH_3 \\ \delta_vH_3
  \end{array}
  \right)
\end{equation}
and obtain a bi-Hamiltonian representation for the next higher
flow
\begin{equation}\label{flow3}
 \left(
  \begin{array}{c}
  u_{t_3} \\ v_{t_3}
  \end{array}
  \right) = J_1 \left(
\begin{array}{c}
\delta_u H_3 \\ \delta_v H_3
\end{array}
\right) = R_1J_1 \left(
\begin{array}{c}
\delta_u H_2 \\ \delta_v H_2
\end{array}
\right) = J_2 \left(
\begin{array}{c}
\delta_u H_2 \\ \delta_v H_2
\end{array}
\right)
\end{equation}
where we have used (\ref{H_3}) and the relation $J_1KJ_1 = R_1J_1
= J_2$, and so on.

\section{Conclusion}

Our starting point was the symplectic and Hamiltonian structure of
the complex Monge-Amp\`ere equation, set into a two-component
evolutionary form. We have calculated all point symmetries of the
$CMA$ system and also, using inverse Noether theorem, Hamiltonians
of the flows for all variational symmetries. These Hamiltonians
yield conservation laws for the $CMA$ flow. We have found two real
$2\times 2$ matrix recursion operators $R_1$ and $R_2$ for
symmetries that commute with the operator ${\cal A}$ of the
symmetry condition and hence map any symmetry of the $CMA$ system
again into a symmetry. The operators $R_1$ and $R_2$ together with
${\cal A}$ form two Lax pairs for the two-component $CMA$ system.
Acting on the first Hamiltonian operator by each recursion
operator, we obtain two new Hamiltonian operators according to
Magri's theorem \cite{magri} and two bi-Hamiltonian
representations of the complex Monge-Amp\`ere equation in the
two-component form. Repeating this action, we could generate an
infinite number of Hamiltonian operators and hence construct a
multi-Hamiltonian representation of the $CMA$ system. We show how
to construct an infinite hierarchy of higher commuting flows
together with the corresponding infinite chain of their
Hamiltonians by using a Hermitian conjugate recursion operator. In
particular, we arrive at three-Hamiltonian representations for
both $CMA$ flow and the first higher flow and bi-Hamiltonian
representations for the zeroth flow and second higher flow. The
results of this paper prove complete integrability of the
(anti-)self-dual gravity in four real dimensions in the sense of
Magri (a multi-Hamiltonian representation).

\section*{Acknowledgements}

One of the authors (MBS) thanks A. A. Malykh for fruitful
discussions. The research of MBS is partly supported by the
research grant from Bogazici University Scientific Research Fund,
research project No. 07B301. One of us (YN) is grateful to Prof.
Dr. Temel Y\i{}lmaz for keeping him alive.

\end{document}